# JCLPCA

## Journal of Cleft Lip Palate and Craniofacial Anomalies

www.jclpca.org

**Inside this issue :**

- In memory and tribute to Kenneth E. Salyer MD: 1936-2020. A great surgeon with a passion for excellence

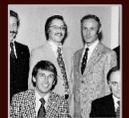
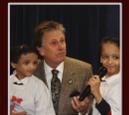

- Building an ecosystem of safe surgery and anesthesia through cleft care

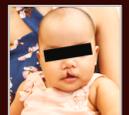
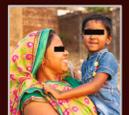

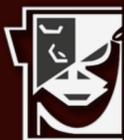

Official Publication of Indian Society of Cleft Lip Palate and Craniofacial Anomalies

# The use of hyaluronic acid in individuals with cleft lip and palate: Literature review

Kelly Fernanda Molena, Lidiane de Castro Pinto[1], Gisele da Silva Dalben[2]

**ABSTRACT**

Since the Resolution 198/2019 of Brazilian Dental Council, which regulates orofacial harmonization as a dental specialty, and the advent of various uses of facial fillers, such as hyaluronic acid (HA), it is possible to perform both esthetic and functional corrections in individuals. Individuals with cleft lip and palate (CLP) present lip irregularities even after orofacial rehabilitation with an interdisciplinary team with several corrective surgeries, interfering with the esthetics, which can cause problems in self-esteem and social insertion. Thus, facial filling is an innovation that, together with dentistry, contributes to the individual's esthetics and well-being. Considering the patient safety and health, more research is progressively being conducted to make such procedures less invasive. This work conducted a literature review on the use of HA as a facial filler to correct lip scars in patients with CLP. By a literature and transverse search in Scientific Electronic Library Online and PubMed databases using specific descriptors, the studies that met the inclusion criteria were selected, from 1990 to 2020. It can be concluded that the use of HA as a facial filling material in the correction of lip scars from reparative surgeries related to CLP has been shown to be effective both for correction of facial asymmetry and to improve the quality of life of patients who used the procedure.

**Key words:** Cleft lip, dentistry, dermal fillers, esthetics, hyaluronic acid

## INTRODUCTION

The Resolution 198/2019 of Brazilian Dental Council acknowledged Orofacial Harmonization (HOF) as a dental specialty and the use of materials as hyaluronic acid (HA) for both esthetic and functional corrections.[1]

Individuals with cleft lip and palate (CLP) can present lip irregularities even after several corrective surgeries, interfering with the esthetics and causing problems in self-esteem and social integration.[2-4]

Thus, the aim of this study was to conduct a literature review regarding the possibility to use HA as a facial filler to correct the lip scar in individuals with CLP. The study comprised a literature review on the use of HA as a facial filler in the correction of lip scars from repair surgeries related to CLP, obtaining the necessary knowledge regarding the effectiveness of using HA as a facial filling material for these scars.

## MATERIALS AND METHODS

A bibliographic cross-sectional search was conducted on articles and electronic media from the following databases: Scientific Electronic Library Online, PubMed (National Center for Biotechnology Information), Google Scholar, as well as a manual search in the reference lists of studies that have been included. The PECOS strategy was used [Table 1] and the research included publications from 1990 to April 2021 in English, Spanish, Italian, French, and Portuguese, with the search terms: "hyaluronic acid" and "cleft lip," and these terms should be present in the title, abstract, or keywords. The following descriptors were used: dermal fillers, cleft lip, and HA. The MeSH

Multiprofessional Resident in Syndromes and craniofacial anomalies, Hospital for Rehabilitation of Craniofacial Anomalies, University of São Paulo, [1]Endodontics Sector, Hospital for Rehabilitation of Craniofacial Anomalies, University of São Paulo, [2]Pediatric and Community Dentistry Sector, Hospital for Rehabilitation of Craniofacial Anomalies, University of São Paulo, Bauru, Brazil

**Address for correspondence:**
DDS. Kelly Fernanda Molena,
Rua Silvio Marchione, 3-20, Vila Nova Cidade universitária, Bauru, São Paulo, 17012-900, Brazil.
E-mail: kelly.molena@usp.br

**Table 1: Population, exposure, comparator, outcomes, study strategy with inclusion criteria of the study**

| PECOS strategy | |
|---|---|
| Element | Description |
| Population | People with cleft lip and palate who have undergone primary correction surgery |
| Exposure | People with cleft lip and palate who underwent primary correction surgery and who used hyaluronic acid on their lip |
| Comparator | People with scars from primary surgeries of cleft lip and palate before being subjected to the use of hyaluronic acid on the lip |
| Outcomes | Improvement of lip asymmetry and quality of life for these people |
| Study design | Case reports between 1990 and April 2021, in full, in English, Spanish, Italian, French, and Portuguese |

PECOS: Population, exposure, comparator, outcomes, study terms were dermal fillers, cleft lip, HA, dentistry, and esthetics. Furthermore, as inclusion criteria, this study included studies with full text available in electronic media and national and international journals. The study excluded theses, thesis chapters, books, book chapters, conference proceedings or reports, technical and scientific reports, and ministerial documents.

The duplicates were removed manually and the selected articles, according to the inclusion and exclusion criteria, were numbered, saved in PDF and organized in a table considering the data:
- Study authors and year of publication
- Study title
- Objective
- Clinical protocol
- Results
- Conclusion of the research.

## RESULTS

Three studies were found in accordance with the inclusion and exclusion criteria on the use of HA to correct scars from primary CLP surgeries. These articles are listed in the table below [Table 2], according to author, year, and publication title.

## LITERATURE REVIEW

CLP is one of the most common craniofacial malformations in mankind, being caused by failure in the union of facial processes in the embryonic period, resulting in a cleft that can extend from the lip to the soft palate, promoting changes in facial morphology, dentition, speech, esthetics, among others.[3] The available findings indicate that CLP affects about 1 in 700 live births. Furthermore, epidemiological and experimental data suggest that it presents a multifactorial origin. Environmental risk factors such as smoking, alcohol, malnutrition, infection, teratogens, and drugs combine with genetic disorders, and the interaction between them is the main causes of malformation.[5]

The treatment of CLP is initiated right after birth and continues until adulthood, requiring the participation of an interdisciplinary team. The morphological rehabilitation of clefts involves lip repair at 3 months of age and palate surgery at around 1 year of age, besides secondary alveolar bone graft performed between 9 and 12 years of age. In addition to primary plastic surgery, rehabilitation requires an interdisciplinary protocol involving different specialties such as speech therapy, maxillofacial surgery, and oral rehabilitation.[4] It is extremely important to standardize the therapeutic procedures, which must be performed by qualified professionals.[6]

The primary surgeries induce the formation of scar tissue in the surgical region, promoting dynamic and static changes that, associated with the cleft, have negative consequences for maxillary growth and development, affecting the entire maxillofacial complex.[7] Over the decades, with advances in surgical techniques, there has been a continuous improvement in the nasolabial appearance of patients with clefts.[8] However, even after several surgical procedures, small deformities, scars, or facial asymmetry may remain, causing a negative impact on facial esthetics, self-esteem, and social inclusion.[2-4,9-11]

Thus, several studies report the perception by professionals, laypeople, parents, and patients about the esthetic result of the treatment of CLP. Lauris *et al.* evaluated the esthetics of the facial profile of children rehabilitated with CLP, comparing the judgment of professionals related and unrelated to the rehabilitation of clefts and laypeople, and the results revealed that professionals related to cleft rehabilitation were kinder and those unrelated to cleft treatment were stricter about facial esthetics than laypeople.[12]

In the studies by Alhayek *et al.* and the professionals were more satisfied than laypeople, corroborating that there are differences in perception between health professionals and laypeople, and the discrepancies between professional groups can be attributed to different treatment modalities and protocols.[13,14] According to Papamanou *et al.* (2012), this can be explained as a result of the education and experience of specialists, which can lead to an assessment of facial

**Table 2: Articles selected in the literature review and organized according to author and year, title, clinical protocol, results, and conclusion**

| Author and year | Publication title | Objective | Clinical protocol | Results | Conclusion |
|---|---|---|---|---|---|
| Stolic D, 2005 | The surgical lips deformity corrected with hyaluronic fillers: A case report | To report a case of use of hyaluronic acid to correct postoperative scars after cleft lip, primary and secondary palate surgery | Local infiltrative 2% lidocaine-epinephrine (lidocaine 40 mg/2 ml+epinephrine 0.025 mg/2 ml) plexus anesthesia, local terminal anesthesia branches of the mandibular nerve and disinfection of the operating field. We used a cross-linked hyaluronic acid (sodium hyaluronate) in the form of a gel concentration of 18.5 mg/g with the addition of antioxidants (mannitol) stylage special lips, VIVACY laboratories, originating in France. For injection is used needle 30 g thickness, length 13 mm factory packed with hyaluronic fillers. They performed a biphasic therapy in the first stage in the zone semimucosis lips is initially carried incision scar tissue. We applied incision fibrous tissue release adhesions. The second stage is placed hyaluronan implant linear retrograde technique and bolus technique whereby the injected 0.33 ml of material, lips were drawn contours and minimum compensation amount of lip volume of 0.5 ml hyaluronic fillers | There was an increase in lip volume, improvement of symmetry and pain and minimal bruising made the patient extremely satisfied with the result | Filling with hyaluronic acid can be used successfully in the treatment of postoperative scars |
| Schweiger ES, 2008 | Successful treatment with injected hyaluronic acid in a patient with lip asymmetry after surgical correction of cleft lip | To present the first report on the use of injectable hyaluronic acid to correct the characteristic lip asymmetry and low volume after surgical repair of a cleft lip | The patient first underwent an intraoral miniblock with 0.5 mL of 1% lidocaine with 1:100,000 epinephrine injected above each canine in the buccal mucosal groove. A quantity of 0.5 mL of HA (restylane, medicis aesthetics, scottsdale, AZ) was then placed into the left mucosal body and vermilion. Additional filler was placed at both areas of dimpled retraction. After the mucosal lip was treated with a goal of 100% correction, an additional 0.2 ml of HA was injected under the left cutaneous upper lip line scar. The patient returned in 10 days for a second application | They obtained a symmetric correction and esthetically pleasing volume augmentation in the affected lip. These results lasted approximately 4 months | The use of HA was able to satisfactorily correct asymmetry and low volume in a surgically repaired lip cleft. For patients who have undergone multiple corrective surgeries, this is a new and less invasive way to improve your cosmetic concerns |
| Franchi G, 2018 | Injections d'acides hyaluroniques au niveau de visages atteints de malformations faciales. Étude préliminaire de l'assouplissement des zones cicatricielles et de l'amélioration esthétique | | A topic anesthetic (such as EMLA® cream) was used and applied 1-2 h before injections. Inhalation of an equimolar mixture of oxygen and nitrous oxide (MEOPA) during injections has been proposed. The same operator applied the injections to all patients in the series, with the following equipment: magnifiers×4; 25 gauge cannulas; 30 ½ inch and 34 gauge needles (length 8 mm); aqueous chlorhexidine, colorless local antiseptic type diluted to 0.2% | | |

| Table 2: Contd... | | | | | |
|---|---|---|---|---|---|
| **Author and year** | **Publication title** | **Objective** | **Clinical protocol** | **Results** | **Conclusion** |
| | | The objective of this study was to evaluate, in addition to the effects of the increase in volume, the effects on the flexibility and elasticity of the scar tissue, of 3 cross-linked hyaluronic acids marketed (doses of 15 mg/ml, 17.5 mg/ml and 20 mg/ml), after reconstructive surgery in patients with congenital or acquired facial anomalies | The three injected implants were marketed, each containing 3 mg/mL of lidocaine hydrochloride and, respectively, had the following dosages of HA (from the least cross-linked to the most cross-linked, that is, from the most fluid to the thickest): 15 mg/ml (Juvederm® VOLBELLA™ with Lidocaine, Allergan, Inc., Irvine, Calif), 17.5 mg/ml (Juvederm® VOLIFT™ with Lidocaine), 20 mg/mL (Juvederm® VOLUMA™ with lidocaine). The linear back-tracing injection was performed in different planes depending on the type of problem to be corrected and the type of HA VOLBELLA™ with lidocaine: Very superficial plane (intradermal, subepidermal, intramucosal, or submucosal) with a 34 gauge needle whose bevel was preferably oriented upward for better control of the amount of gel administered; VOLIFT™ with lidocaine: intermediate plane (deep intradermal, subdermal or submucosal) with 34 gauge needle; VOLUMA™ with Lidocaine: Deep plane (intra-adipose or bone contact) with ½ inch 30 gauge needle or 25 gauge cannula | The cross-linked hyaluronic acid fillers offered very subtle cosmetic results and complemented the surgery with a high level of patient satisfaction. When injected into fibrosis, the first session increased smoothness and elasticity; the second session increased the volume. Cross-linked hyaluronic acid fillers fill in the sunken areas and improve the softness and elasticity of scar tissues | In addition to its well-known space-filling function as a side effect, the authors demonstrate that cross-linked hyaluronic acid fillers improve the softness and elasticity of scar tissue. Many experimental studies support this, showing that cross-linked hyaluronic acid stimulates the production of various components of the extracellular matrix, including dermal collagen and elastin |

HA: Hyaluronic acid, MEOPA: Equimolar mixture of oxygen and nitrous oxide, AZ: Arizona esthetics from a different perspective compared to laypeople. From the perspective of parents and patients, low satisfaction with esthetics was correlated with the increase in self-reported influence of the cleft on the social activity and professional life of patients, as well as their high expectations with the results.[14] These findings highlight the observed negative influence of the cleft on the patient's social activity and professional life and underline the need for the highest quality of surgical outcome for this group of patients.[15-19]

Meyer-Marcotty (2011) analyzed the facial perception of patients with unilateral cleft compared with Class III orthognathic patients and concluded that there was a greater degree of asymmetry in patients with cleft nose and lower lip, even when surgery was performed on the young patient. Although Class III patients were classified as less attractive compared to controls, patients with clefts were classified as significantly less attractive, showing that not only the extent of this asymmetry influences the attractiveness but also its location.[20]

With this in mind, orofacial fillers as HA are versatile agents for filling the lips and perioral region and can act as adjuvants for esthetic rehabilitation and improvement of quality of life in these patients. It plays several roles in the formation and repair of tissues, collagen synthesis, elasticity, and support of the skin. Since this is a reversible procedure, it is very safe for cosmetic and dental procedures.[21,22]

Schweiger *et al*. reported the use of HA filling for correction of lip asymmetry in an adult patient with unilateral CLP, which presented lip asymmetry resulting from primary lip and palate surgeries. An amount of 0.5 ml of HA was placed on the body and border of the left lip vermilion with excellent cosmetic results and durability of around 4 months.[23]

In 2015, Stolic *et al*. applied HA in a patient with scarred lip resulting from the primary surgeries of CLP, but in two stages, release of fibrous tissue through an incision and shortly after the insertion of the acid hyaluronic obtaining volume and lip symmetry for this patient.[24]

Kandhari *et al*. reported the use of HA as an alternative for correcting congenital, acquired, and postsurgical lip asymmetries in three clinical cases, in which there was significant improvement in lip symmetry and also in the lip and perioral profile of these patients.[25]

Franchi *et al*., besides evaluating the increase in lip volume, also analyzed the effects of HA on the flexibility and elasticity of the scar tissue and were able to verify that, when there was a hardened scar, the first session of injections improved flexibility and elasticity, but it provided a moderate volume since the tissues were not deformable. The second session had an essentially volumizing effect since the tissues were softened and could be more easily expanded. Thus, due to its viscoelastic properties, the application of HA in patients with CLP improved the asymmetry and also the lip flexibility and elasticity.[26]

Accordingly, Hussain *et al.* and Khan *et al.* reported in their work that the use of HA improved the appearance of dystrophic scars. Thus, the use of HA to correct lip scar in patients with CLP, based on the current literature, is an option for the esthetic rehabilitation of these patients, in addition to improving their self-esteem and social insertion.[27,28]

## DISCUSSION

CLP, being one of the most common craniofacial malformations in mankind, presents esthetic sequels represented by scars or facial asymmetry as a result of treatment with reparative surgeries, causing a negative impact on facial esthetics, self-esteem, and social inclusion.[2-4,9-11] Throughout their lives, patients with CLP undergo various surgical and reconstructive procedures and, at the end of these, a scar or lip asymmetry may persist.

Facial fillers are an important innovation in the field of dentistry, because, by this procedure, it is possible to provide support to improve, correct, and prevent the damage to the facial tissues.[29] Among these, HA is one of the best fillers currently used; a member of the family of glycosaminoglycans, it is present in the fundamental intercellular substance of animal tissues and in the capsules of certain bacteria, presenting a high capacity to bind to water, making it an interesting substance to add volume to the skin.[30] HA as a filling biomaterial has practical applications, good safety margin, besides immediate and lasting esthetic effects. Its biocompatibility and relatively simple application technique have made it a frequent choice for facial volumization, being effective for correcting asymmetries, refining scars and skin flexibility, thus enabling corrections of scars resulting from reparative surgeries, and improving the quality of life of these patients, as a way to achieve the esthetic harmony not achieved by surgeries.[21,23,26,27]

Eventually, mild swelling may occur after application, which usually disappears within 24 h. However, in most cases, the patient is normally able to return to the routine simply by applying cold compresses and taking analgesics and/or anti-inflammatories prescribed by the dentist.[29,31] In case of exacerbated volumization, as well as in the treatment of nodules, granulomas, or necrosis, it is possible to apply hyaluronidase aiming at resorption of the product. Furthermore, after application, it is recommended to perform a vigorous massage on the spot to homogenize the area, and the professional must have full knowledge of anatomy to avoid applications in blood vessels.[21,31]

For individuals with CLP, HA is shown as a less invasive treatment option, since this is a nonsurgical procedure, with quick and easy application and more affordable. Evaluating HA in the correction of lip asymmetry in patients with scars from primary surgeries to correct CLP, it was noted that this material can offer, by a minimally invasive method, considerable esthetic improvement and lip flexibility and elasticity, thus also improving the lip function.[23,26]

Other authors have reported its use for correcting scars in other regions, such as the cheek and forearm, with considerable improvement in their appearance.[27,28] The volume and concentration depend on the application site, and the procedure and time needed for reapplication vary between patients, with an average of 3 months to 1 year, thus it is fundamental to perform individual planning for each patient.[21,28,31]

Thus, with Resolution 198/2019, which recognizes orofacial HOF as a dental specialty, it appears as a great opportunity in the rehabilitation of individuals with CLP, since a properly qualified dentist can continue the treatment to meet the patient's esthetic expectations and thus provide the integrity of rehabilitation, which also includes the psychological aspects.

## CONCLUSION

Based on findings of the literature of the past 20 years, the use of HA as a facial filling material in the correction of lip scars from reparative surgeries related to CLP has been shown to be effective both for correction of facial asymmetry and to improve the quality of life of patients submitted to this procedure.

### Financial support and sponsorship
Nil.

### Conflicts of interest
There are no conflicts of interest.